\documentclass[12pt]{article}

\usepackage{graphicx}

\def\stackunder#1#2{\mathrel{\mathop{#2}\limits_{#1}}}%

\begin{document}

\author{Georges Ripka\footnote{Email: ripka@spht.saclay.cea.fr} \\
ECT*, villa Tambosi, I-38050\ Villazzano (Trento) Italy\\
and\\
Service de Physique Th\'{e}orique, Centre d'Etudes de Saclay, \\
F-91191\ Gif-sur-Yvette, Cedex}
\title{Quantum fluctuations of the quark condensate.}
\date{March 15, 2000}
\maketitle

\begin{abstract}
The quantum fluctuations of the quark condensate are studied in a Nambu
Jona-Lasinio model.\ Two Lorenz invariant regularizations are considered: a
sharp 4-momentum cut-off and a soft gaussian regulator. The quantum
fluctuations of the quark condensate are found to be large although chiral
symmetry is not restored. Instabilities of the ground state appear when the
system is probed by a source term proportional to the squared quark
condensate. The instabilities are traced to unphysical poles introduced by
the regulator and their effect is greatly enhanced when a sharp cut-off is
used.
\end{abstract}

\section{Introduction.}

We study the effect of quantum fluctuations of the quark condensate on the
physical vacuum. We use an $SU\left( 2\right) _f$ Nambu Jona-Lasinio model
in the chiral limit. Two ultra-violet regularizations are considered:

\begin{itemize}
\item  regularization using a sharp cut-off, in which the quark propagators
are set to zero when $k_\mu ^2>\Lambda ^2$, where $k_\mu $ is the euclidean
4-momentum of the quark propagator;

\item  regularization using a smooth gaussian regulator which will be
described below.
\end{itemize}

We introduce a current quark mass $m$ as a source term $m\bar{\psi}\psi $ to
calculate the quark condensate $\left\langle \bar{\psi}\psi \right\rangle $.
We introduce another source term $\frac 12j\left( \bar{\psi}\Gamma _a\psi
\right) ^2$ which allows us to calculate the expectation value $\left\langle
\left( \bar{\psi}\Gamma _a\psi \right) ^2\right\rangle $ of the \emph{squared%
} quark condensate. We shall however maintain $j$ at a finite value in order
to study the response of the system to a deformation induced by this source
term. We include the $1/N_c$ effects due to the quantum fluctuations of the
meson fields. In the quark language, this means that we include the exchange
(Fock) term as well as the (RPA) ring diagrams.

Although the quantum fluctuations of the meson fields do not restore chiral
symmetry, we do find surprisingly large fluctuations of the quark
condensate. We also find that the effective potential is very sensitive to
the shape of the regulator. Apparent instabilities appear which we show to
be artefacts sharp cut-of used in conjunction with the relatively low values
of the cut-offs, used in chiral quark models.

The quantum fluctuations of the meson fields in the Nambu Jona-Lasinio model
have been studied for several years \cite{Lemmer95}, \cite{Ripka96b}, \cite
{Broniowski96}, \cite{Oertel00}. In these studies, the quark and meson loops
were regularized with different cut-offs. Since both the quark and meson
loops diverge, their relative contribution could be adjusted at will by a
proper choice of the cut-offs thereby making it impossible to estimate the
importance of the quantum fluctuations of the meson fields.

In this study we consider the Nambu Jona-Lasinio model to be a quark model
in the sense that all physical processes can be expressed in term of Feynman
graphs involving only quark propagators. The meson fields which are
introduced in the process of bosonization are mere intermediate quantities
introduced to calculate the partition function. When the quark propagators
are regularized, a single cut-off regularizes both the quark and meson
loops. No further regularization is required for higher order loops. This
approach has also been adopted in Ref.\cite{Huefner98} for example, with the
exception that a 3-momentum cut-off was used.\ We shall show that results
obtained with 3 and 4-momentum cut-offs can differ even qualitatively. We
show that regularization with a 4-momentum cut-off introduces a non-locality
which makes the ground state energy unbounded from below, or, equivalently,
which makes the model acausal. We relate this instability to the unphysical
poles of the quark propagator which are introduced by the regulator.

We consider the regularization of a model to be a physical phenomenon and
not simply a way to be rid of infinities wherever they turn up. Because of
this we regularize the model action from the outset before calculating the
loop integrals and we shall see that this makes a significant difference
with the common practice of regularizing the infinities of the loop
integrals which are deduced from an unregularized action. We derive all
quantities in terms of a regulator which is a function of the squared
euclidean 4-momentum $k^2$.

The paper is composed of three parts.\ In the first part we set the
notation, we explain how the calculation was performed and we define the
relevant range of the model parameters. The second part is devoted to a
discussion of the quantum fluctuations of the quark condensate.\ The last
part discusses the instabilities which are displayed by the response of the
system, subject to constraint proportional to the squared quark condensate.

\section{The model euclidean action.}

The model is defined by the euclidean action: 
\begin{equation}
I_m\left( q,\bar{q}\right) =\left\langle \bar{q}\left| -i\partial _\mu
\gamma _\mu +rmr\right| q\right\rangle -\frac{g^2}{2N_c}\int d_4x\left(
\left\langle \bar{q}\left| r\right| x\right\rangle \Gamma _a\left\langle
x\left| r\right| q\right\rangle \right) ^2  \label{nonlocact}
\end{equation}
It involves a quark field $q\left( x\right) \equiv \left\langle x\left|
q\right. \right\rangle $. The euclidean Dirac matrices are $\gamma _\mu
=\gamma ^\mu =\left( i\beta ,\vec{\gamma}\right) $. The model includes a
regulator $r$ which is assumed to be diagonal in $k$-space: $\left\langle
k\left| r\right| k^{\prime }\right\rangle =\delta _{kk^{\prime }}r\left(
k\right) $.\ We consider both a gaussian regulator: 
\begin{equation}
r\left( k\right) =e^{-\frac{k^2}{2\Lambda ^2}}  \label{gaussian}
\end{equation}
and a sharp cut-off: 
\begin{equation}
r\left( k\right) =1\;\;if\;\;k^2<\Lambda ^2\quad \quad \quad r\left(
k\right) =0\;\;if\;\;k^2>\Lambda ^2  \label{sharp}
\end{equation}
The use of the the sharp cut-off is tantamount to the calculation of Feynman
graphs in which the quark propagators are set to zero when their euclidean
4-momentum $k_\mu ^2$ exceeds the cut-off $\Lambda ^2$. The bra-ket notation
for the Dirac fields is: 
\[
\left\langle \bar{q}\left| -i\partial _\mu \gamma _\mu +m\right|
q\right\rangle \equiv \int d_4x\,\bar{q}\left( x\right) \left( -i\partial
_\mu \gamma _\mu +m\right) q\left( x\right) 
\]
\begin{equation}
\left\langle x\left| r\right| q\right\rangle =\int d_4y\,\left\langle
x\left| r\right| y\right\rangle \,q\left( y\right) \quad \quad \left\langle 
\bar{q}\left| r\right| x\right\rangle =\int d_4y\,\,\bar{q}\left( y\right)
\,\left\langle y\left| r\right| x\right\rangle \quad \quad
\end{equation}
We use the current quark mass $m$ to evaluate a regularized quark condensate
and that is why the current quark mass $m$ is multiplied by the regulator.
The $\Gamma _a=\left( 1,i\gamma _5\vec{\tau}\right) $ are the generators of
chiral rotations. The number of flavors is $N_f$ and there are $N_f^2$
generators $\Gamma _a$. We assumed that the coupling constant $\frac{g^2}{N_c%
}$ is inversely proportional to $N_c$.

The partition function $W$ is given by the euclidean path integral\footnote{%
Pedantically, the partition function is $Z=e^{-W}$. In the following we
shall call $W$ the partition function without fear of confusion.}: 
\begin{equation}
e^{-W\left( m\right) }=\int D\left( q\right) D\left( \bar{q}\right)
e^{-I\left( q,\bar{q}\right) }  \label{partaa}
\end{equation}

At zero temperature and for an infinite translationally invariant system, $%
W=\Omega \varepsilon $ where $\Omega $ is the space-time volume $\Omega
=\int d_4x\,1$ and where $\varepsilon $ is the energy density $\varepsilon
=\frac EV$, which is the energy per unit volume of the system in its ground
state.

The model is regularized in the same way as the effective quark model which
has been derived from a study of the propagation of quarks in an instanton
liquid \cite{Shuryak82},\cite{Diakonov86} in which case both the shape of
the regulator and the value of the cut-off are derived. However, the model
action (\ref{nonlocact}) we are using is not exactly the same as that
derived from the instanton liquid \cite{Diakonov96}. The model described by
the action (\ref{nonlocact}) has been actively investigated in both the
soliton \cite{Ripka98} and meson sectors \cite{Birse98} \cite
{Broniowski99,Broniowski00}.

We now adopt a notation which adds considerable transparency to the
manipulations made below.\ We define an interaction $V$ by its matrix
element: 
\begin{equation}
\left\langle xa\left| V\right| yb\right\rangle =\left\langle yb\left|
V\right| xq\right\rangle =-\delta _{ab}\delta \left( x-y\right) \frac{g^2}{%
N_c}\quad \quad \quad \left\langle xa\left| V^{-1}\right| yb\right\rangle
=-\delta _{ab}\delta \left( x-y\right) \frac{N_c}{g^2}  \label{vmat}
\end{equation}
We define a \emph{delocalized} quark field $\psi \left( x\right) $ and the
corresponding bilinear forms $\bar{\psi}\Gamma _a\psi $: 
\begin{equation}
\psi \left( x\right) =\int d_4x\,\left\langle x\left| r\right|
y\right\rangle \,q\left( y\right) \quad \quad \quad \bar{\psi}\left(
x\right) \Gamma _a\psi \left( x\right) =\int d_4y\,d_4z\,\bar{q}\left(
y\right) \left\langle y\left| r\right| x\right\rangle \Gamma _a\left\langle
x\left| r\right| z\right\rangle q\left( z\right)
\end{equation}
In this notation, the action (\ref{nonlocact}) takes the form: 
\begin{equation}
I_m\left( q,\bar{q}\right) =\left\langle \bar{q}\left| -i\partial _\mu
\gamma _\mu +rmr\right| q\right\rangle +\frac 12\left( \bar{\psi}\Gamma \psi
\right) V\left( \bar{\psi}\Gamma \psi \right)  \label{nonlocact2}
\end{equation}
The chiral limit is defined to be $m\rightarrow 0$.

\section{The constrained system and the effective potential.}

The current quark mass $m$ serves as a source term to calculate the
normalized quark condensate $\left\langle \bar{\psi}\psi \right\rangle $.\
In order to calculate the quantum fluctuations of the quark condensate, we
introduce an extra source term: 
\begin{equation}
\frac 12j\left( \bar{\psi}\Gamma _a\psi \right) ^2=\frac 12j\int
d_4x\,\left[ \left( \bar{\psi}\psi \right) ^2+\left( \bar{\psi}i\gamma
_5\tau _a\psi \right) ^2\right]
\end{equation}
which acts as a constraint on the system and which allows us to calculate
the expectation value $\left\langle \left( \bar{\psi}\Gamma _a\psi \right)
^2\right\rangle $ of the \emph{squared} condensate. We prefer to work with
the squared condensate $\left( \bar{\psi}\Gamma _a\psi \right) ^2$ rather
than with $\left( \bar{\psi}\psi \right) ^2$ because $\bar{\psi}\psi $ is
only one of the four components of the chiral 4-vector $\bar{\psi}\Gamma
_a\psi $. If the system chooses to vibrate in the direction $\bar{\psi}\psi $
defined by the ground state, it will do so because the amplitudes of the
vibrations in the four directions are independent. But there is no reason to
prevent the system to vibrate in the other three directions.

The constraint is introduced into the action (\ref{nonlocact2}) and we
define \emph{the constrained system} by the partition function: 
\begin{equation}
e^{-W\left( j,m\right) }=\int D\left( q\right) D\left( \bar{q}\right)
e^{-I_m\left( q,\bar{q}\right) +\frac 12\left( \bar{\psi}\Gamma \psi \right)
j\left( \bar{\psi}\Gamma \psi \right) }=\int D\left( q\right) D\left( \bar{q}%
\right) e^{-I_{j,m}\left( q,\bar{q}\right) }  \label{wjqq}
\end{equation}
where: 
\begin{equation}
I_{j,m}\left( q,\bar{q}\right) =\left\langle \bar{q}\left| -i\partial _\mu
\gamma _\mu +rmr\right| q\right\rangle +\frac 12\left( \bar{\psi}\Gamma \psi
\right) \left( V-j\right) \left( \bar{\psi}\Gamma \psi \right)
\end{equation}
is the action of the constrained system.

The ground state expectation value $\left\langle \left( \bar{\psi}\Gamma
_a\psi \right) ^2\right\rangle $ of the squared condensate in the
constrained system is given by: 
\begin{equation}
\frac 12\Omega \left\langle \left( \bar{\psi}\Gamma _a\psi \right)
^2\right\rangle =-\left. \frac{\partial W\left( j,m\right) }{\partial j}%
\right| _{m=0}  \label{wjcon}
\end{equation}
and the quark condensate is given by the expression: 
\begin{equation}
\Omega \left\langle \bar{\psi}\psi \right\rangle =\left. \frac{\partial
W\left( j,m\right) }{\partial m}\right| _{m=0}  \label{psibarpsi}
\end{equation}
where $\Omega $ is the space-time volume in euclidean space.

The constrained system, described by the partition function $W\left(
j,m\right) $ is not the same as the system described by the partition
function $W\left( m\right) $ because it has an additional potential energy
equal to $-\frac 12j\left\langle \left( \bar{\psi}\Gamma _a\psi \right)
^2\right\rangle =j\frac{\partial W\left( j,m\right) }{\partial j}$. This is
why, in the presence of the constraint, the energy density of the system is
defined in terms of the \emph{effective potential}: 
\begin{equation}
\Gamma =W\left( j,m\right) +\frac 12j\left\langle \left( \bar{\psi}\Gamma
_a\psi \right) ^2\right\rangle =\left. W\left( j,m\right) -j\frac{\partial
W\left( j,m\right) }{\partial j}\right| _{m=0}  \label{effact}
\end{equation}
Viewed as a function of $j$ (for a fixed $m$), a stationary point of the
action occurs when: 
\begin{equation}
\frac{\partial W\left( j,m\right) }{\partial j}=-j\frac{\partial ^2W\left(
j,m\right) }{\partial j^2}
\end{equation}
so that the effective action is stationary when $j=0$, that is, in the
absence of the constraint. This is true whatever approximation we use to
calculate $W\left( j,m\right) $.

The effective potential allows us to map out the energy of the system as it
deforms under the effect of the constraint $\frac 12j\left( \bar{\psi}\Gamma
_a\psi \right) ^2$. The choice of the constraint used to probe the energy
surface of the system is, of course, arbitrary.\ Our choice is justified by
the fact that, as we shall see in section \ref{sec:effpotclass}, the system
offers a relatively soft response to the constraint and in some cases it
actually displays an instability of the ground state.

\section{Bosonization in terms of local fields.}

\label{sec:bosonization}

Bosonization is simply a convenient way to calculate the partition function $%
W\left( j,m\right) $. We introduce \emph{local} auxiliary fields $\varphi
_a\left( x\right) $ and we consider the new euclidean action: 
\begin{equation}
I_{j,m}\left( q,\bar{q},\varphi \right) =\left\langle \bar{q}\left|
-i\partial _\mu \gamma _\mu +rmr+r\varphi _a\Gamma _ar\right| q\right\rangle
-\frac 12\varphi \left( V-j\right) ^{-1}\varphi  \label{iqqphiact}
\end{equation}
Consider the partition function $W^{\prime }\left( j,m\right) $ defined by
the path integral: 
\begin{equation}
e^{-W^{\prime }\left( j,m\right) }=\int D\left( \varphi \right) D\left(
q\right) D\left( \bar{q}\right) e^{-I_{j,m}\left( q,\bar{q},\varphi \right) }
\label{wdash}
\end{equation}
The integral over $\varphi $ yields: 
\begin{equation}
\int D\left( \varphi \right) e^{-\frac 12\left( \left( \bar{\psi}\Gamma \psi
\right) -\varphi \left( V-j\right) ^{-1}\right) \left( V-j\right) \left(
\left( \bar{\psi}\Gamma \psi \right) -\left( V-j\right) ^{-1}\varphi \right)
}=e^{\frac 12tr\ln \left( V-j\right) }  \label{trlnv}
\end{equation}
where $tr$ denotes a trace in $\left( x,a\right) $ space: 
\begin{equation}
tr\,1=\Omega N_f^2
\end{equation}
(because there are $N_f^2$ generators $\Gamma _a$). From (\ref{wdash}) and (%
\ref{trlnv}) we deduce a relation between the partition functions $W^{\prime
}\left( j,m\right) $ and $W\left( j,m\right) $: 
\begin{equation}
W\left( j,m\right) =W^{\prime }\left( j,m\right) +\frac 12tr\ln \left(
V-j\right)
\end{equation}
If we integrate out the quarks in the path integral (\ref{wdash}), we get
the partition function $W\left( j,m\right) $ in the form: 
\begin{equation}
e^{-W\left( j,m\right) }=\int D\left( \varphi \right) e^{-I_{j,m}\left(
\varphi \right) -\frac 12tr\ln \left( V-j\right) }  \label{wj}
\end{equation}
where $I_{j,m}\left( \varphi \right) $ is the so-called ''bosonized
action'': 
\begin{equation}
I_{j,m}\left( \varphi \right) =-Tr\ln \left( -i\partial _\mu \gamma _\mu
+rmr+r\varphi _a\Gamma _ar\right) -\frac 12\varphi \left( V-j\right)
^{-1}\varphi  \label{ij2}
\end{equation}
The trace $Tr$ is over the variables (space-time $x$, Dirac indices, flavor
and color) which define the quark field: 
\begin{equation}
Tr\,\,1=4N_cN_f\Omega
\end{equation}
It is convenient to shift the constituent quark mass to the interaction term
by writing $m+\varphi _a\Gamma _a\equiv \varphi _a^{\prime }\Gamma _a$. The
bosonized action (\ref{ij2}) becomes: 
\begin{equation}
I_{j,m}\left( \varphi \right) =-Tr\ln \left( -i\partial _\mu \gamma _\mu
+r\varphi _a\Gamma _ar\right) -\frac 12\left( \varphi -m\right) \left(
V-j\right) ^{-1}\left( \varphi -m\right)  \label{ij}
\end{equation}
where dropped the prime on $\varphi $. The partition function is then given
by the expression (\ref{wj}) in which $I_{j,m}\left( \varphi \right) $ is
the action (\ref{ij}). In the expression (\ref{ij}), $m$ stands for the
vector $m_a$ with components $m_a=\left( m,0,0,0\right) $.

It is usual to calculate the effective potential from the texbook formula 
\cite{Itzykson80}: 
\begin{equation}
\Gamma \left( \varphi \right) =I_m\left( \varphi \right) +\frac 12tr\ln 
\frac{\partial ^2I_m\left( \varphi \right) }{\partial \varphi \partial
\varphi }
\end{equation}
obtained from a Legendre transform which relates a source term to the
expectation value of the field. We do not use this formalism because, in our
case, the operator $\frac{\partial ^2I_m\left( \varphi \right) }{\partial
\varphi \partial \varphi }$ has negative eigenvalues.\ This is easily seen
by considering cuts in the Mexican hat shaped $I_m\left( \varphi \right) $.

\section{The classical approximation to the constrained system.}

\label{sec:classpot}

The classical approximation consists in approximating the path integral (\ref
{wj}) as: 
\begin{equation}
W\left( j,m\right) =I_{j,m}\left( \varphi _j\right)
\end{equation}
where $\varphi _j$ is a stationary point of the action $I_{j,m}\left(
\varphi \right) $, defined in (\ref{ij}). In the classical approximation, we
also neglect the constant $\frac 12tr\ln \left( V-j\right) $ which we will
include in section \ref{sec:fluctuations} together with the contribution of
the field fluctuations. The term ''classical'' will be used throughout
although, at the quark level, it includes the quark loop. As it turns out,
it is the classical approximation \emph{to the constrained system}
determines the response of the system to the constraint and the quantum
fluctuations of the fields are only small corrections. The classical
approximation is the leading order contribution in $N_c$ and, in the quark
representation, it corresponds to the Hartree approximation.

\subsection{The gap equation and the relation between $j$ and $M$.}

\label{sec:gapequation}

A stationary point of the action (\ref{ij}) occurs for $\varphi _a=\left(
M,0,0,0\right) $ where $M$ is the solution of the so-called gap equation: 
\begin{equation}
\left( V-j\right) ^{-1}=-\frac M{M-m}\frac 1{2\Omega }Tr\frac{r^4}{-\partial
_\mu ^2+r^4M^2}\equiv -4N_cN_f\frac M{M-m}g_M  \label{gap}
\end{equation}
where $g_M=g_M\left( q=0\right) $ and $g_M\left( q\right) $ is the function (%
\ref{gmq}), defined in appendix \ref{app:mesonprop}. We shall denote by $%
I_{j,m}\left( M\right) $ the action at the stationary point. An explicit
expression is given in appendix \ref{app:classact}.

Let $M_0$ be the solution of the gap equation in the absence of a constraint 
$\left( j=0\right) $: 
\begin{equation}
V^{-1}=-4N_cN_f\frac{M_0}{M_0-m}\,g_{M_0}  \label{gmv}
\end{equation}
This equation relates $M_0$ to the interaction strength $V$. The minus sign
indicates that the interaction $V$ is attractive. Otherwise chiral symmetry
would not be spontaneously broken and $\varphi _a=0$ would be the only
stationary point of the action.

We can calculate $j$ from the equation: 
\begin{equation}
j=\frac 1{4N_cN_f}\left( \frac 1{g_M}-\frac 1{g_{M_0}}\right) -\frac
m{4N_cN_f}\left( \frac 1{Mg_M}-\frac 1{M_0g_{M_0}}\right)  \label{jmj}
\end{equation}

In practice, we start by choosing a value of $M_0$, which determines $V$. We
then choose a value of $M$ which determines $j$. We can easily derive the
relation: 
\begin{equation}
\frac{\delta M^2}{\delta j}=-4N_fN_cg_M^2\left( \frac{dg_M}{dM^2}\right)
^{-1}\quad \quad \left( m=0\right)  \label{dmmdj}
\end{equation}
which shows that $M$ is a monotonically increasing function of $j$ so that
it makes little difference if we plot the effective potential as a function
of $j$ or $M$. The choice of $M$ is more transparent.

\subsection{The classical estimates of $\left\langle \bar{\psi}\psi
\right\rangle $ and $\left\langle \left( \bar{\psi}\Gamma _a\psi \right)
^2\right\rangle $ in the chiral limit.}

In the classical approximation, the quark condensate (\ref{psibarpsi}) is
given by the expression: 
\begin{equation}
\Omega \left\langle \bar{\psi}\psi \right\rangle _{class}=\left. \frac{%
\partial I_{j,m}\left( M\right) }{\partial m}\right| _{m=0}=\Omega \left(
V-j\right) ^{-1}M=-4\Omega N_cN_fMg_M  \label{classcond}
\end{equation}
The classical approximation to the squared quark condensate (\ref{wjcon})
is: 
\begin{equation}
\frac 12\Omega \left\langle \left( \bar{\psi}\Gamma _a\psi \right)
^2\right\rangle _{class}=-\frac{dI_{j,m}\left( M^2\right) }{dj}=\frac
12\Omega M^2\left( 4N_cN_fg_M\right) ^2  \label{conclass}
\end{equation}
where we used the fact that, for a given value of $j$, $M$ is a stationary
point of the action $I_{j,m}\left( M\right) $ .

It follows that, in the classical approximation, the \emph{fluctuation} of
the quark condensate is zero: 
\begin{equation}
\left\langle \left( \bar{\psi}\Gamma _a\psi \right) ^2\right\rangle
_{class}=\left\langle \bar{\psi}\psi \right\rangle _{class}^2
\end{equation}
as expected. The quantum fluctuations of the quark condensate are due to the
quantum fluctuations of the $\varphi $ fields and they are introduced in
section \ref{sec:fluctuations}.

The gap equation (\ref{gap}) relates $M$ to the classical quark condensate
and $M$ is simply a Hartree insertion:

\begin{equation}
\begin{tabular}{cc}
$M=\left( V-j\right) \left\langle \bar{\psi}\psi \right\rangle _{class}$ & $%
\includegraphics[width=3cm,height=1cm]{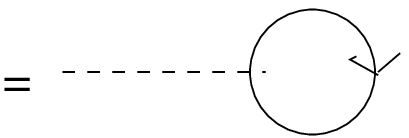}$%
\end{tabular}
\label{graph12}
\end{equation}

\subsection{The classical effective potential in the chiral limit.}

Because $M$ is a stationary point of the action $I_{j,m}\left( M\right) $,
the classical approximation to the effective potential (\ref{effact}) is: 
\begin{equation}
\Gamma _{class}=I_{j,m}\left( M\right) -j\frac{\partial I_{j,m}\left(
M\right) }{\partial j}=I_{j,m}\left( M\right) +j\frac 12\Omega \left(
V-j\right) ^{-2}M^2  \label{classpot2}
\end{equation}
Using (\ref{gap}) and (\ref{jmj}), the classical effective potential can
also be expressed as the following function of $M$: 
\begin{equation}
\Gamma _{class}\left( M\right) =I_{j,m}\left( M\right) +\frac 12\Omega
\left( 4N_cN_f\right) M^2g_M\left( 1-\frac{g_M}{g_{M_0}}\right) \quad \quad
\quad \left( m=0\right)  \label{classpot}
\end{equation}
An explicit expression for $I_{j,m}\left( M\right) $ is given in appendix 
\ref{app:classact}.

\section{Inclusion of the quantum fluctuations of the fields $\varphi $.}

\label{sec:fluctuations}

A saddle point evaluation of the partition function (\ref{wj}) yields: 
\begin{equation}
W\left( j,m\right) =I_{j,m}\left( \varphi _j\right) +\left. \frac 12tr\ln 
\frac{\delta ^2I_{j,m}\left( \varphi \right) }{\delta \varphi \delta \varphi 
}\right| _{\varphi =\varphi _j}+\frac 12tr\ln \left( V-j\right)
\label{potloop}
\end{equation}
where $\varphi _j$ is the stationary point of $I_{j,m}\left( \varphi \right) 
$, which is determined by the gap equation (\ref{gap}).

\subsection{Calculation of $W\left( j,m\right) $.}

To calculate the second term of (\ref{potloop}), we need to evaluate the
inverse meson propagator matrix: 
\begin{equation}
K_{ab}^{-1}\left( x,y\right) =\frac{\delta ^2I_{j,m}\left( \varphi \right) }{%
\delta \varphi _a\left( x\right) \delta \varphi _b\left( y\right) }
\end{equation}
at the stationary point of $I_{j,m}\left( \varphi \right) $. From the action
(\ref{ij}), we see that: 
\begin{equation}
K^{-1}=\Pi -\left( V-j\right) ^{-1}  \label{kpivj}
\end{equation}
where $\Pi $ is the polarization function (Lindhardt function): 
\[
\Pi _{ab}\left( x,y\right) =-\frac{\delta ^2}{\delta \varphi _a\left(
x\right) \delta \varphi _b\left( y\right) }Tr\ln \left( -i\partial _\mu
\gamma _\mu +rmr+r\varphi _a\Gamma _ar\right) 
\]
\begin{equation}
=Tr\,\,\frac 1{-i\partial _\mu \gamma _\mu +rmr+r\varphi _a\Gamma _ar}\left|
x\right\rangle \Gamma _a\left\langle x\right| \frac 1{-i\partial _\mu \gamma
_\mu +rmr+r\varphi _a\Gamma _ar}\left| y\right\rangle \Gamma _b\left\langle
y\right|  \label{piab}
\end{equation}
The partition function (\ref{potloop}) becomes: 
\begin{equation}
W\left( j,m\right) =I_{j,m}\left( \varphi _j\right) +\left. \frac 12tr\ln
\left( 1-\Pi \left( V-j\right) \right) \right| _{\varphi =\varphi _j}
\label{wjmkp}
\end{equation}

The matrices $\Pi $ and $V$, and therefore $K$, are diagonal in momentum
space and in the flavor indices: 
\[
\left\langle qa\left| K^{-1}\right| q^{\prime }b\right\rangle =\delta
_{ab}\delta _{qq^{\prime }}K_a^{-1}\left( q\right) \quad \quad \quad
\left\langle qa\left| \Pi \right| q^{\prime }b\right\rangle =\delta
_{ab}\delta _{qq^{\prime }}\Pi _a\left( q\right) 
\]
In particular: 
\[
\Pi _{a=0}\left( q\right) \equiv \Pi _S\left( q\right) =4N_cN_f\left( \frac
12q^2f_M^{22}\left( q\right) +M^2\left( f_M^{26}\left( q\right)
+f_M^{44}\left( q\right) \right) -g_M\left( q\right) \right) 
\]
\begin{equation}
\Pi _{a=1,2,3}\left( q\right) \equiv \Pi _P\left( q\right) =4N_cN_f\left(
\frac 12q^2f_M^{22}\left( q\right) +M^2\left( f_M^{26}\left( q\right)
-f_M^{44}\left( q\right) \right) -g_M\left( q\right) \right)  \label{k2abis}
\end{equation}
where the functions $f_M^{np}\left( q\right) $ and $g_M\left( q\right) $ are
defined by the expressions (\ref{fnpq}) and (\ref{gmq}) of appendix \ref
{app:mesonprop}. The regulator $r$ ensures that the polarization functions $%
\Pi \left( q\right) $ vanish when $q>>2\Lambda $.

If we use the gap equation (\ref{gap}) to express $\left( V-j\right) ^{-1}$
in terms of $g_M\equiv g_M\left( q=0\right) $, we obtain analogous
expressions for the inverse meson propagators: 
\[
K_{a=0}^{-1}\left( q\right) \equiv K_S^{-1}\left( q\right) =4N_cN_f\left(
\frac 12q^2f_M^{22}\left( q\right) +M^2\left( f_M^{26}\left( q\right)
+f_M^{44}\left( q\right) \right) -g_M\left( q\right) +\frac M{M-m}g_M\right) 
\]
\begin{equation}
K_{a=1,2,3}^{-1}\left( q\right) \equiv K_P^{-1}\left( q\right)
=4N_cN_f\left( \frac 12q^2f_M^{22}\left( q\right) +M^2\left( f_M^{26}\left(
q\right) -f_M^{44}\left( q\right) \right) -g_M\left( q\right) +\frac
M{M-m}g_M\right)  \label{kaq2}
\end{equation}
The pion remains a Goldstone boson even in the constrained system because,
in the chiral limit, $K_{a=1,2,3}^{-1}\left( q\right) \stackunder{%
q\rightarrow 0}{\rightarrow }0$. This is an important feature of the choice
we have made of the constraint which does not break the chiral symmetry of
the lagrangian.

The partition function (\ref{wjmkp}) is then: 
\[
W\left( j,m\right) =I_{j,m}\left( M^2\right) 
\]
\[
+\frac 12\sum_q\ln \left( 1+\frac{M-m}M\frac 1{g_M}\left( \frac
12q^2f_M^{22}\left( q\right) +M^2\left( f_M^{26}\left( q\right)
+f_M^{44}\left( q\right) \right) -g_M\left( q\right) \right) \right) 
\]
\begin{equation}
+\frac{N_f^2-1}2\sum_q\ln \left( 1+\frac{M-m}M\frac 1{g_M}\left( \frac
12q^2f_M^{22}\left( q\right) +M^2\left( f_M^{26}\left( q\right)
-f_M^{44}\left( q\right) \right) -g_M\left( q\right) \right) \right)
\label{wj3}
\end{equation}
It is because we took the trouble to keep track of the term $\frac 12tr\ln
\left( V-j\right) $ that the sums over $q$ converge. Another way to write $%
W\left( j,m\right) $ is: 
\begin{equation}
W\left( j,m\right) =I_{j,m}\left( M\right) +\frac 12\sum_{qa}\ln \frac{%
K_a^{-1}\left( q\right) }{4N_cN_fg_M}\frac{M-m}M  \label{wj4}
\end{equation}

\subsection{The Fock term (exchange energy) and the ring diagrams.}

Consider the expression (\ref{wjmkp}) of $W\left( j,m\right) $. Using the
explicit form (\ref{ij}) of the action $I_{j,m}\left( M\right) $ as well as
the relation (\ref{graph12}), we can write the partition function $W\left(
j,m\right) $, in the chiral limit, as follows: 
\begin{equation}
W\left( j,m\right) =-Tr\ln \left( -i\partial _\mu \gamma _\mu +r^2M\right)
-\frac 12\left\langle \bar{\psi}\psi \right\rangle \left( V-j\right)
\left\langle \bar{\psi}\psi \right\rangle +\left. \frac 12tr\ln \left( 1-\Pi
\left( V-j\right) \right) \right| _{\varphi =\varphi _j}  \label{feynw}
\end{equation}
We can express the second term, together with the expansion of the
logarithm, in terms of Feynman graphs. The second term of (\ref{feynw}) is
the direct (Hartree) energy:

\begin{equation}
\begin{tabular}{cc}
$\frac 12\left\langle \bar{\psi}\psi \right\rangle \left( V-j\right)
\left\langle \bar{\psi}\psi \right\rangle =$ & $%
\includegraphics[width=4cm,height=1cm]{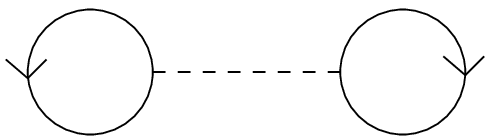}$%
\end{tabular}
\label{graph1}
\end{equation}
The expansion of the logarithm in (\ref{feynw}) generates the exchange
(Fock) energy as well as the ring diagrams:

\[
\frac 12tr\ln \left( 1-\Pi \left( V-j\right) \right) 
\]
\[
=-\frac 12tr\Pi \left( V-j\right) +\frac 14tr\Pi \left( V-j\right) \Pi
\left( V-j\right) -\frac 16tr\Pi \left( V-j\right) \Pi \left( V-j\right) \Pi
\left( V-j\right) +... 
\]
\begin{equation}
\includegraphics[width=15cm,height=4cm]{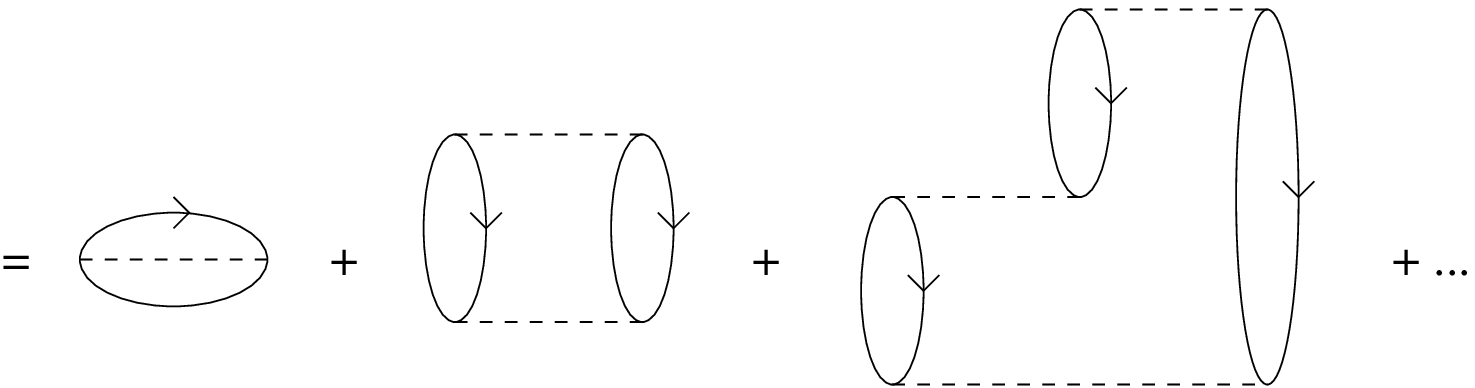}  \label{graph2}
\end{equation}
The terms of (\ref{graph2}) are the next to leading order in $N_c$. The
first term, which is the exchange (Fock) term, is special in that it does
not induce any correlations in the quark wavefunction. It is easy to see
that the exchange term is more sensitive than the ring diagrams to the high
momenta running in the meson loop.\ As a result, when a sharp cut-off is
used, it is the exchange term and not the ring diagrams which dominates the
next to leading order contribution (\ref{graph2}).

\subsection{Expressions for the quark condensates and the quark condensates
in the chiral limit.}

The ground state expectation value of the squared condensate (\ref{wjcon})
can be calculated from the expression: 
\begin{equation}
\frac 12\Omega \left\langle \left( \bar{\psi}\Gamma _a\psi \right)
^2\right\rangle =\left. \frac{\partial I_{j,m}\left( M\right) }{\partial j}%
\right| _{m=0}-\frac{\delta M^2}{\delta j}\frac \delta {\delta M^2}\left(
\frac 12\sum_{qa}\ln \frac{K_a^{-1}\left( q\right) }{4N_cN_fg_M}\right)
\label{sqcon}
\end{equation}
where $\frac{\delta M^2}{\delta j}$ is given by (\ref{dmmdj}). The
contribution of the exchange term to the squared condensate is:

\begin{equation}
\frac 12\Omega \left\langle \left( \bar{\psi}\Gamma _a\psi \right)
^2\right\rangle _{exch}=\frac{\delta M^2}{\delta j}\frac \delta {\delta
M^2}\left( \frac 12\left( V-j\right) \sum_{qa}\Pi _a\left( q\right) \right)
\end{equation}

The quark condensate (\ref{psibarpsi}) can be calculated with the help of (%
\ref{wjmkp}): 
\begin{equation}
\Omega \left\langle \bar{\psi}\psi \right\rangle =\frac{\partial W\left(
j,m\right) }{\partial m}=\frac \partial {\partial m}\left( I_{j,m}\left(
M\right) +\left. \frac 12tr\ln \left( 1-\Pi \left( V-j\right) \right)
\right| _M\right)
\end{equation}
We calculate the derivative $\frac{\partial W\left( j,m\right) }{\partial m}$
keeping $j$ constant.\ However, we must remember that $M$ depends on both $j$
and $m$ and therefore a contribution arises from the change in $M$ when $m$
is varied because, in contrast to $I_{j,m}$, $W\left( j,m\right) $ is not
stationary with respect to variations of $M$. When $m\rightarrow m+\delta m$%
, $\left( V-j\right) $ remains constant and: 
\begin{equation}
\delta \frac 12tr\ln \left( 1-\left( V-j\right) \Pi \right) =-\frac 12tr%
\frac{V-j}{1-\left( V-j\right) \Pi }\delta \Pi =\frac 12trK\delta \Pi =\frac
12\frac{\delta M^2}{\delta m}trK\frac{\delta \Pi }{\delta M^2}
\end{equation}
The quark condensate is thus: 
\begin{equation}
\left\langle \bar{\psi}\psi \right\rangle =\left\langle \bar{\psi}\psi
\right\rangle _{class}+\frac{\delta M^2}{\delta m}\frac 12\sum_{qa}K_a\left(
q\right) \frac{\delta \Pi _a\left( q\right) }{\delta M^2}  \label{qcond}
\end{equation}
From the gap equation (\ref{gap}) we see that $\frac{\delta M^2}{\delta m}=-%
\frac{g_M}M\left( \frac{dg_M}{dM^2}\right) ^{-1}$.

The contribution of the exchange term to the quark condensate is: 
\begin{equation}
\Omega \left\langle \bar{\psi}\psi \right\rangle _{exch}=-\left( V-j\right) 
\frac{\delta M^2}{\delta m}\frac 1{2\Omega }\sum_{qa}\frac{\delta \Pi
_a\left( q\right) }{\delta M^2}  \label{qcoondexch}
\end{equation}

The effective potential (\ref{effact}) can be calculated from the expression 
\begin{equation}
\Gamma =W\left( j,m\right) +j\frac 12\left\langle \left( \bar{\psi}\Gamma
_a\psi \right) ^2\right\rangle  \label{effpot}
\end{equation}
where $j$ is given by (\ref{jmj}) and $\frac 12\left\langle \left( \bar{\psi}%
\Gamma _a\psi \right) ^2\right\rangle $ by (\ref{sqcon}).

\section{The relevant range of parameters.}

We consider now the parameters of the model. We calculate all quantities in
units of the cut-off $\Lambda $ which appears in the regulators (\ref
{gaussian}${}$) and (\ref{sharp}).\ This is convenient because, for example,
the effective potential is proportional to $\Lambda ^4$ but otherwise it
depends only on the ratios $M_0/\Lambda $ and $M/\Lambda $. Thus, the
stability of the system depends on the single parameter $M_0/\Lambda $ and,
of course on the shape of the regulator. The ratio $M_0/\Lambda $ is
therefore the key parameter to consider and we must determine the physically
meaningful range of values for $M_0/\Lambda $.

In the vicinity of $q=0$, the inverse pion propagator determines the value
of $f_\pi $.\ In the chiral limit, we have: 
\begin{equation}
K_P^{-1}\left( q\right) \approx Z_P\,\,q^2\quad \quad \quad f_\pi =M_0\sqrt{%
Z_P}  \label{fpi}
\end{equation}
and $Z_P$ is the residue of the pion propagator at the pole $q=0$. For a
given value of $M_0/\Lambda $, the value of $f_\pi $ depends on the value of 
$M_0$. Tables \ref{table:fpisharp} and \ref{table:fpigauss} give the values
of $Z_P$ and $f_\pi $ for various values of $\frac{M_0}\Lambda $ and for two
values of $M_0$, namely $M_0=300\,\,MeV$ and $M_0=400\,\,MeV$. For $%
M_0=300\;MeV$, the observed value $f_\pi =93\;MeV$ is fitted with $\frac{M_0}%
\Lambda \approx 0.4$.\ For $M_0=\,400\,MeV$, higher values $\frac{M_0}%
\Lambda \approx 0.6$ and $\frac{M_0}\Lambda \approx 0.7$ are required
respectively, when a sharp cut-off and a gaussian regulator are used.
Soliton calculations require $M_0$ to lie between $300$ and $400\;MeV$ \cite
{Ripka98}.\ Higher values of $M_0$ and therefore of $M_0/\Lambda $ have also
been considered in order to push the unphysical $q\bar{q}$ continuum well
above the $\rho $ mass of $770\,\;MeV$.\ In Ref.\cite{Oertel00} for example,
the value $M_0/\Lambda =0.74$ is used. In order to cover the full range of
physically meaningful parameters, we perform our calculations from $%
M_0/\Lambda =0.2$, which means a relatively high value of the cut-off, up to 
$M_0/\Lambda =0.8$, which means a low value of the cut-off.

\begin{table}[tbp] \centering%
$
\begin{tabular}{|c|c|c|c|c|}
\hline
$\frac{M_0}\Lambda $ & $Z_S$ & $Z_P$ & $
\begin{array}{c}
f_\pi \,\,\left( MeV\right) \\ 
{\rm for }M_0=300\,\,MeV
\end{array}
$ & $
\begin{array}{c}
f_\pi \,\,\left( MeV\right) \\ 
{\rm for }M_0=400\,\,MeV
\end{array}
$ \\ \hline
0.2 & 0.171 & 0.189 & 130 & 173 \\ \hline
0.4 & 0.0825 & 0.0964 & 93.2 & 124 \\ \hline
0.6 & 0.0433 & 0.0533 & 69.7 & 92.4 \\ \hline
0.8 & 0.0239 & 0.0307 & 52.6 & 70.1 \\ \hline
1.0 & 0.0138 & 0.0183 & 40.7 & 54.2 \\ \hline
\end{tabular}
$%
\caption{Values of $Z_S$, $Z_P$ and $f_\pi $, calculated with a sharp cut-off, for various values of $\frac{M_0}\Lambda $ and for two values of $M_0$.
\label{table:fpisharp}}%
\end{table}%

\begin{table}[tbp] \centering%
$
\begin{tabular}{|c|c|c|c|c|}
\hline
$\frac{M_0}\Lambda $ & $Z_S$ & $Z_P$ & $
\begin{array}{c}
f_\pi \,\,\left( MeV\right) \\ 
{\rm for }M_0=300\,\,MeV
\end{array}
$ & $
\begin{array}{c}
f_\pi \,\,\left( MeV\right) \\ 
{\rm for }M_0=400\,\,MeV
\end{array}
$ \\ \hline
0.2 & 0.0766 & 0.1621 & 121 & 161 \\ \hline
0.4 & 0.0305 & 0.0950 & 92.5 & 123 \\ \hline
0.6 & 0.0155 & 0.0659 & 77.0 & 103 \\ \hline
0.8 & 0.0089 & 0.0498 & 67.0 & 89.3 \\ \hline
1.0 & 0.0056 & 0.0395 & 59.7 & 79.6 \\ \hline
\end{tabular}
$%
\caption{Values of $Z_S$, $Z_P$ and $f_\pi $, calculated with a gaussian regulator, for various values of $\frac{M_0}\Lambda $ and for two values of $M_0$. 
\label{table:fpigauss}}%
\end{table}%

One crucial point here is that the relevant range of parameters (typically $%
0.4<M_0/\Lambda <0.8$) corresponds to uncomfortably \emph{low} values of the
cut-off. Most field theoretic methods applied to statistical mechanics and
particle physics have been developed to systems in which $M_0<<\Lambda $.
Considerable errors can be (and have been) made by applying methods and
concepts, borrowed from the study of systems in which $M_0<<\Lambda $, and
applied to systems in which the cut-off $\Lambda $ is of the same order of
magnitude as $M_0$. Our calculations focus on several problems which one
encounters when the cut-off is not much larger than the calculated
observables. This is the regime applicable to low-energy hadronic physics
and it is not an artefact of the Nambu Jona-Lasinio\ model. For example, low
energy effective theories derived from an instanton liquid \cite
{Diakonov86,Diakonov96} yield values of $M_0/\Lambda $ of the order of $0.4$.

As an example, consider the inverse $\sigma $-meson propagator at $q=0$ in
the chiral limit.\ It is given by (\ref{kaq2}): 
\begin{equation}
K_S^{-1}\left( q=0\right) =8N_cN_fM^2f_M\equiv 4M^2Z_S
\end{equation}
where $f_M$ is the function $f_M^{np}\left( q\right) $, defined in (\ref
{fnpq}), and taken at $q=0$, where it is independent of $n$ and $p$. This
definition of $Z_S$ is quite arbitrary except for the fact that, for large
values of the cut-off, that is, when $\frac{M_0}\Lambda \rightarrow 0$, we
have $Z_S=Z_P$ and the Nambu Jona-Lasinio model reduces to a linear sigma
model. The values of $Z_S$ are also listed in the tables \ref{table:fpisharp}
and \ref{table:fpigauss}. We see that, with a sharp cut-off, $Z_S$ and $Z_P$
differ by about 30\% in the relevant parameter range $\frac{M_0}\Lambda
=0.4\,-\,0.8$. With a gaussian cut-off the equality $Z_S=Z_P$ is not even
approximately obtained. The fact that $Z_S\neq Z_P$ contradicts most, if not
all previously reported calculations of the meson propagators, when they are
derived from an unregularized action with loop integrals subsequently
regularized. If we had proceeded this way, the loop integrals $%
f_M^{np}\left( q\right) $ would have been independent of $n$ and $p$, and
the function $g_M\left( q\right) $ would have become independent of $q$.\
Instead of the expression (\ref{kaq2}), the meson propagators would have
been equal to the usually quoted expressions (in the chiral limit): 
\begin{equation}
K_S^{-1}\left( q\right) =4N_cN_f\frac 12\left( q^2+4M^2\right) f_M\left(
q\right) \quad \quad K_P^{-1}\left( q\right) =4N_cN_f\frac 12q^2f_M\left(
q\right)
\end{equation}
and $Z_S$ would equal $Z_P$. The tables \ref{table:fpisharp} and \ref
{table:fpigauss} show that considerable errors can be introduced if the
regularization is not specified in the action from the outset and adhered
to. Of course, these errors would be small if the cut-off were large, that
is, if $M_0/\Lambda $ were very small.\ However, in the applications of the
Nambu Jona-Lasinio model to low energy hadronic physics, it is not.

The recently claimed instability of the Nambu Jona-Lasinio model, heralded
by Kleinert et al.\cite{Kleinert99}, is based on a reduction of the Nambu
Jona-Lasinio model to a non-linear $\sigma $-model.\ Although the arguments
presented above raise doubts as to the validity of this reduction, and such
doubts have also been voiced elsewhere \cite{Babaev00}, we shall show in
section \ref{sec:effpotclass} that instabilities do indeed arise but that
they are artefacts of the sharp cut-off regularization when the cut-off is
close to $M_0$.

\section{The quark condensate of the unconstrained system.}

In this section we consider the quark condensate (\ref{qcond}) of the
unconstrained system $\left( j=0\right) $. Various contributions to the
quark condensate are listed in table \ref{table:condensate} for various
values of $M_0/\Lambda $. They are expressed in units of $\Lambda ^3$. We
see that, throughout the range of relevant parameters, the quadratic
fluctuations of the fields do not alter significantly the quark condensate.\
They show no sign of restoring chiral symmetry. These results agree with
those found in Ref.\cite{Broniowski00}. A finer analysis would show that the
negative contributions of the pion field are due to the exchange term.
Although the field fluctuations do not alter significantly the ground state
expectation value of the quark condensate, we shall see in the next section
that they do cause an appreciable \emph{quantum fluctuation} of the
condensate.

\begin{table}[tbp] \centering%
\begin{tabular}{|c|c|c|c|c|}
\hline
$\frac{M_0}\Lambda $ & $\left\langle \bar{\psi}\psi \right\rangle _{class}$
& $\left\langle \bar{\psi}\psi \right\rangle _{\,\sigma \;loop}$ & $%
\left\langle \bar{\psi}\psi \right\rangle _{\pi \;loop}$ & $\left\langle 
\bar{\psi}\psi \right\rangle _{total}$ \\ \hline
\begin{tabular}{c}
0.4 \\ 
sharp
\end{tabular}
& -0.0392 & 0.0018 & -0.0030 & -0.0404 \\ \hline
\begin{tabular}{c}
0.6 \\ 
sharp
\end{tabular}
& -0.0446 & 0.0004 & 0.0003 & -0.0439 \\ \hline
\begin{tabular}{c}
0.8 \\ 
sharp
\end{tabular}
& -0.0451 & -0.0001 & 0.0020 & -0.0432 \\ \hline
\begin{tabular}{c}
0.4 \\ 
gauss
\end{tabular}
& -0.0218 & 0.0024 & -0.0026 & -0.0220 \\ \hline
\begin{tabular}{c}
0.6 \\ 
gauss
\end{tabular}
& -0.0275 & 0.0016 & -0.0011 & -0.0270 \\ \hline
\begin{tabular}{c}
0.8 \\ 
gauss
\end{tabular}
& -0.0315 & 0.0012 & -0.0001 & --0.0304 \\ \hline
\end{tabular}
\caption{The classical quark condensate and the contributions of the quadratic fluctuations of the $\sigma $ and $\pi $ fields.\ The last column sums all the contributions. The condensates are expressed in units of $\Lambda ^3$.
\label{table:condensate}}%
\end{table}%

\section{The quantum fluctuations of the quark condensate in the
unconstrained system.}

The expressions (\ref{qcond}) and (\ref{sqcon}), which give the quark
condensate $\left\langle \bar{\psi}\psi \right\rangle $ and the squared
condensate $\left\langle \left( \bar{\psi}\Gamma _a\psi \right)
^2\right\rangle $, allow us to calculate the \emph{quantum} \emph{fluctuation%
} of the quark condensate $\Delta _{\bar{\psi}\psi }$: 
\begin{equation}
\Delta _{\bar{\psi}\psi }=\sqrt{\left\langle \left( \bar{\psi}\Gamma _a\psi
\right) ^2\right\rangle -\left\langle \bar{\psi}\psi \right\rangle ^2}
\end{equation}
The values are listed in table \ref{table:condfluc} for various values of $%
M_0/\Lambda $. The fluctuations are due to the exchange and ring diagrams
and they vanish in the classical approximation.\ We see that, relative to
the quark condensate, they are quite large: about 50\% when a sharp cut-off
is used and between 70\% and 80\% when a gaussian regulator is used. We also
see that the quark condensates are more sensitive to the shape of the
regulator than $f_\pi $.\ This is because $f_\pi $ would diverge
logarithmically with a large cut-off, whereas the quark condensates would
have a quadratic dependence on the cut-off. This is also the reason why,
once $f_\pi $ is fixed, larger condensates are obtained with a sharp cut-off
than with a gaussian regulator.

\begin{table}[tbp] \centering%
\begin{tabular}{|c|c|c|c|c|}
\hline
$\frac{M_0}\Lambda $ & $\left\langle \bar{\psi}\psi \right\rangle $ & $%
\left\langle \left( \bar{\psi}\Gamma _a\psi \right) ^2\right\rangle $ & $%
\Delta _{\bar{\psi}\psi }$ & $\frac{\Delta _{\bar{\psi}\psi }}{\left\langle 
\bar{\psi}\psi \right\rangle }$ \\ \hline
\begin{tabular}{c}
0.2 \\ 
sharp
\end{tabular}
& -0.0300 & $11.69\times 10^{-4}$ & 0.0164 & 0.55 \\ \hline
\begin{tabular}{c}
0.4 \\ 
sharp
\end{tabular}
& -0.0405 & $20.05\times 10^{-4}$ & 0.0192 & 0.47 \\ \hline
\begin{tabular}{c}
0.6 \\ 
sharp
\end{tabular}
& -0.0438 & $23.32\times 10^{-4}$ & 0.0200 & 0.46 \\ \hline
\begin{tabular}{c}
0.8 \\ 
sharp
\end{tabular}
& -0.0432 & $22.56\times 10^{-4}$ & 0.0196 & 0.45 \\ \hline
\begin{tabular}{c}
0.2 \\ 
gaussian
\end{tabular}
& -0.0145 & $4.38\times 10^{-4}$ & 0.0151 & 1.03 \\ \hline
\begin{tabular}{c}
0.4 \\ 
gaussian
\end{tabular}
& -0.0220 & $7.75\times 10^{-4}$ & 0.017 & 0.78 \\ \hline
\begin{tabular}{c}
0.6 \\ 
gaussian
\end{tabular}
& -0.0270 & $10.67\times 10^{-4}$ & 0.018 & 0.68 \\ \hline
\begin{tabular}{c}
0.8 \\ 
gaussian
\end{tabular}
& -0.0304 & $12.97\times 10^{-4}$ & 0.019 & 0.63 \\ \hline
\end{tabular}
\caption{The quark condensate $\left\langle \bar{\psi}\psi \right\rangle $, in units of $\Lambda ^3$, the squared condensate $\left\langle \left( \bar{\psi}\Gamma _a\psi \right) ^2\right\rangle $, in units of $\Lambda ^6$, the fluctuation of the condensate $\Delta$, in units of $\Lambda ^3$ and its relative value $\Delta _{\bar{\psi}\psi }/\left\langle \bar{\psi}\psi \right\rangle $. 
\label{table:condfluc}}%
\end{table}%

\section{The effective potential in the chiral limit.}

\label{sec:effpotclass}

\begin{figure}[h]
\centerline{\resizebox*{12cm}{10cm}{\includegraphics{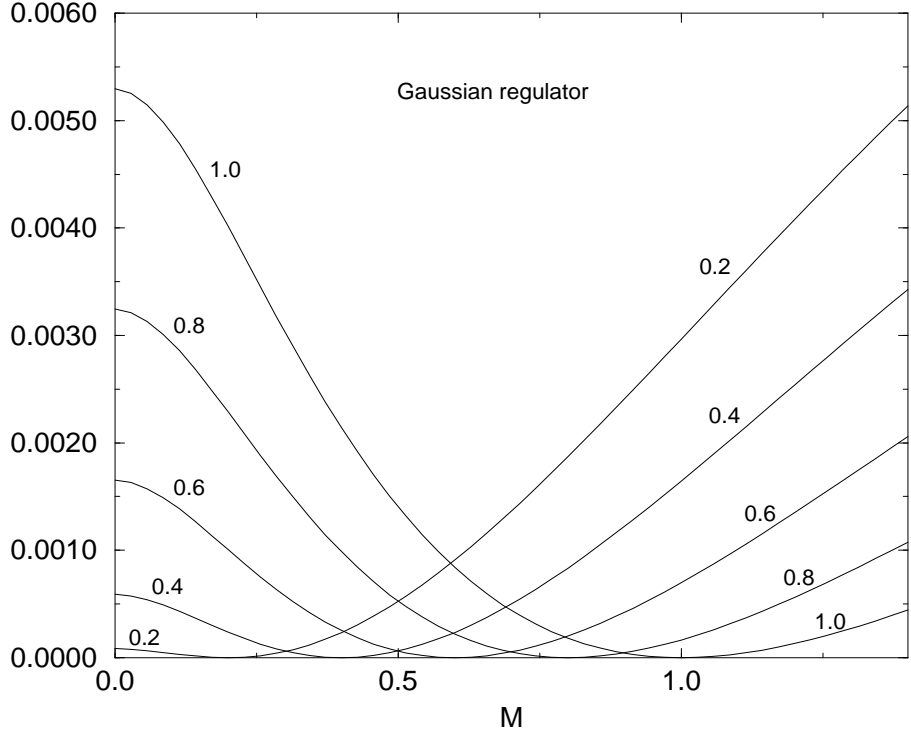}} }
\caption{The classical effective potential calculated with a gaussian
regulator and plotted as a function of $M$, for various indicated values of $%
M_0/\Lambda $. The effective potential is expressed in units of $\Lambda ^4$.
}
\label{epsfig52}
\end{figure}

Figure \ref{epsfig52} shows the classical effective potential (\ref{classpot}%
) in the chiral limit, as a function of $\frac M\Lambda $, calculated with a
gaussian regulator, for various values of $\frac{M_0}\Lambda $. The minimum
occurs at $M=M_0$ and in all this work, we define the zero of energy to be
equal to the minimum of the classical action at the point $M=M_0$. These
curves map out the energy surface of the system while it is being deformed
by the constraint $\frac 12j\left( \bar{\psi}\Gamma _a\psi \right) ^2$. As
stressed in section \ref{sec:gapequation}, it makes no difference whether we
plot the effective potential as a function of $j$, or of $M$. Plotted as a
function of $M$, the curves are easier to understand.

\begin{figure}[h]
\centerline{\resizebox*{12cm}{10cm}{\includegraphics{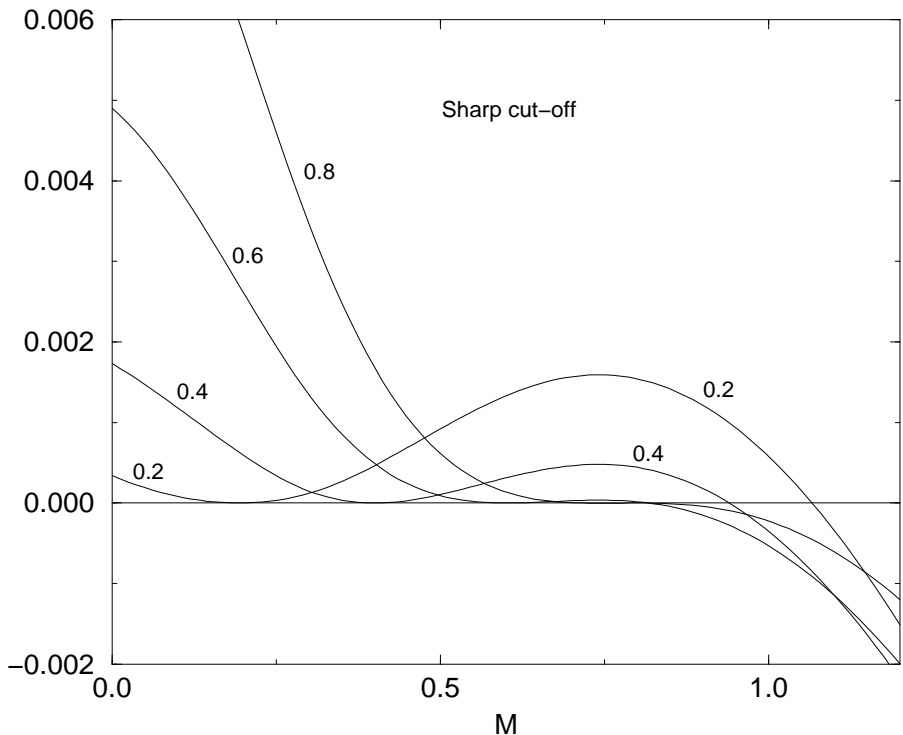}} }
\caption{The classical effective potential, calculated with a sharp 4-momentum cut-off, and plotted as a function of $M$ for various indicated values of $M_0/\Lambda $. The effective potential is expressed in units of $\Lambda ^4$.}
\label{epsfig57}
\end{figure}

Figure \ref{epsfig57} shows shows the classical effective potential (\ref
{classpot}) calculated with a sharp cut-off. The code can be checked against
analytic expressions in this case. We see that for increasing values of $%
\frac{M_0}\Lambda $, that is, for decreasing values of the cut-off, the
minimum of the effective potential at $M=M_0$ becomes increasingly shallower
and that it disappears altogether at the critical value $\frac{M_0}\Lambda
\geq 0.\,742\quad $which can be evaluated analytically. When a gaussian
regulator is used, the onset of the instability occurs at the much higher
value $\frac{M_0}\Lambda \geq 2.93$ which was evaluated numerically.

The system appears to display an instability with respect to perturbations
caused by the constraint $\frac 12j\left( \bar{\psi}\Gamma _a\psi \right) ^2$%
. Furthermore, the energy of the system does not seem bounded from below. It
goes without saying that the classical action (as opposed to the classical
effective potential) displays a minimum at $M=M_0$ for all values of $%
M_0/\Lambda $.

\begin{figure}[h]
\centerline{\resizebox*{12cm}{10cm}{\includegraphics{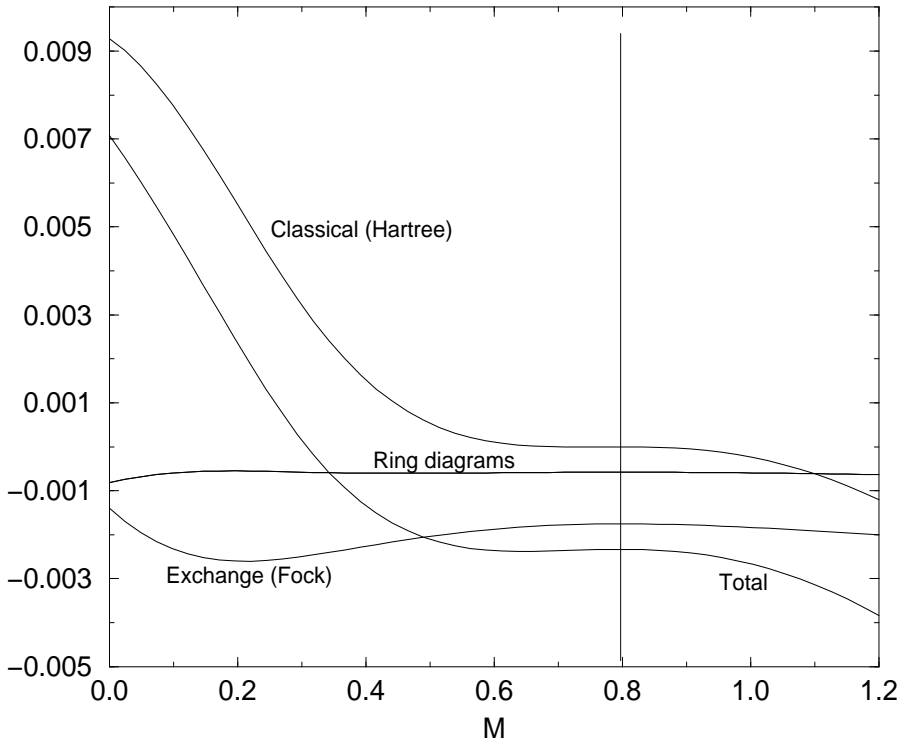}} }
\caption{Various contributions to the effective potential calculated with a sharp cut-off with a value $M_0/\Lambda =0.8$. The energies are measured relative to the classical action at $M=M_0$. }
\label{epsfig47}
\end{figure}

Let us take a closer look at this apparent instability.\ It is not an
artefact of the classical approximation. Figure \ref{epsfig47} shows the
various contributions to the effective potential when the $1/N_c$ field
fluctuations are included, using a sharp cut-off with $M_0/\Lambda =0.8$. We
see that the field fluctuations lower the energy but that they do not
significantly change the shape of the effective potential, so that the
instability remains. We also see that the exchange (Fock) term dominates the 
$1/N_c$ corrections because it is more sensitive than the ring diagrams to
the high momenta running in the meson loop. These conclusions remain valid
for smaller values of $M_0/\Lambda $.

\begin{figure}[h]
\centerline{\resizebox*{12cm}{10cm}{\includegraphics{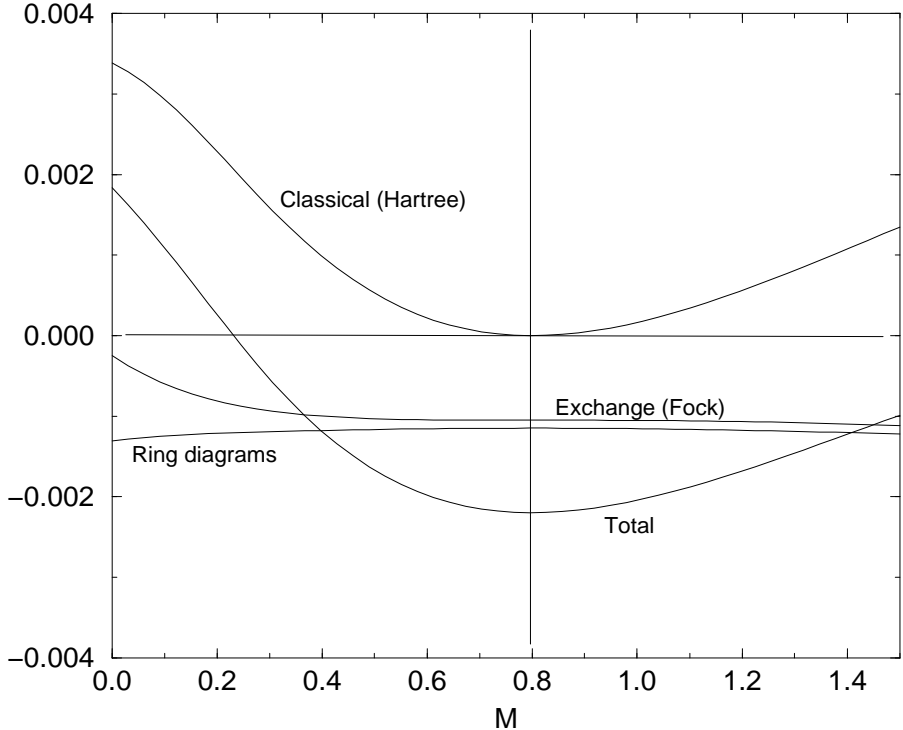}} }
\caption{Various contributions to the effective potential calculated with a gaussian regulator and with $M_0/\Lambda =0.8$. The energies are measured relative to the classical action at $M=M_0$.}
\label{epsfig49}
\end{figure}

By way of comparison, figure \ref{epsfig49} shows the contributions of the $%
1/N_c$ corrections when a gaussian regulator is used. Although the effect of
the meson fluctuations is somewhat larger, the stability of the system is
not modified. We see however that the ring diagrams make a somewhat larger
contribution than the exchange (Fock) term, because the gaussian regulator
reduces the effect of the high momenta running in the meson loop. The same
conclusions can be reached for different values of $\frac{M_0}\Lambda $. As
mentioned above, the instability also occurs when a gaussian cut-off is
used, but at much higher values of $M_0/\Lambda \geq 2.93$.\ For such high
values, the cut-off is too small to be physically meaningful. With a
gaussian regulator and in the relevant range of parameters $0.4<M_0/\Lambda
<0.8$, one needs to probe the system with values as high as $M/\Lambda >4$
before it becomes apparent that the energy is not bounded from below.

\begin{figure}[h]
\centerline{\resizebox*{12cm}{10cm}{\includegraphics{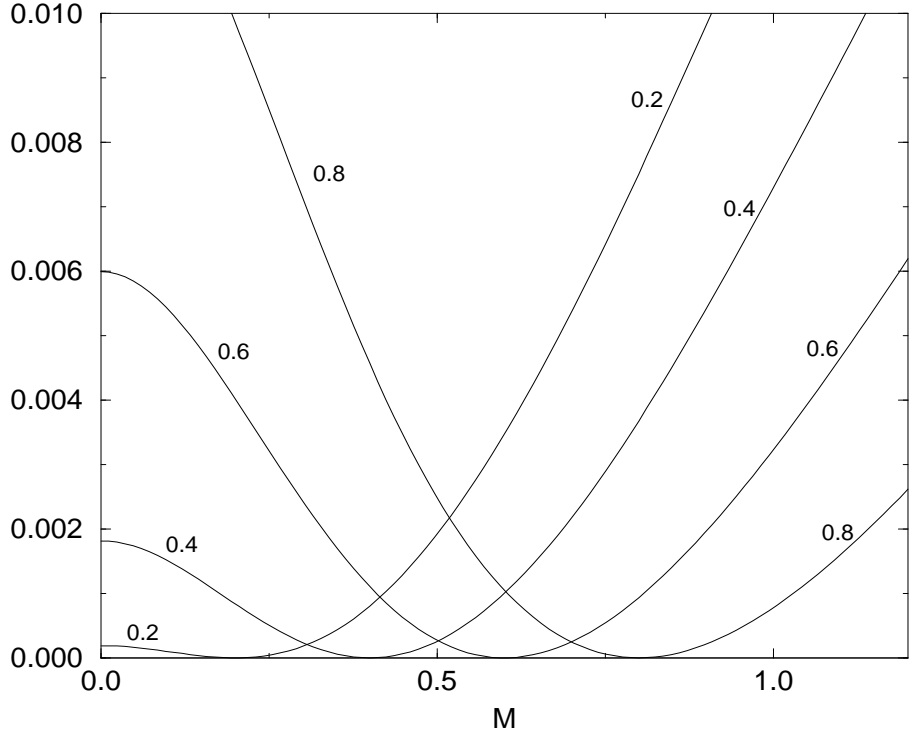}} }
\caption{The classical effective potential calculated with a sharp 3-momentum cut-off, for various indicated values of $M_0/\Lambda $.\ No instability appears.}
\label{epsfig58}
\end{figure}

A clue concerning the nature of the instability can be obtained by
considering the effective potential obtained with a sharp 3-momentum
regularisation.\ In this regularization, the trace of the quark loop is
calculated by integrating the energy variable from $-\infty $ to $+\infty $,
and by limiting the 3-momentum by the condition $\left| \vec{k}\right|
<\Lambda $. This regularization is tantamount to a limitation of the quantum
mechanical Hilbert space available to the quarks. Figure \ref{epsfig58}
shows that the effective potential, calculated with a sharp 3-momentum
cut-off, behaves as expected and that it does not display the instability.

\section{Unphysical poles of the quark propagator.}

The fact that the effective potential, calculated with a sharp 4-momentum
cut-off, displays an instability that does not occur when a sharp 3-momentum
cut-off is used, can be understood as an effect of the unphysical poles of
the quark propagator which are introduced by the 4-momentum regulator. For
constant fields, the quark propagator can be written in the form: 
\begin{equation}
\frac 1{k_\mu \gamma _\mu +r^2M}=\frac{-k_\mu \gamma _\mu +r_k^2M}{\omega ^2+%
\vec{k}^2+r_k^4M^2}\quad \quad \quad \left( k_\mu =\left( \omega ,\vec{k}%
\right) \right)
\end{equation}
When a 3-momentum regularisation is used, we have $r_k^2=1$ when $\left| 
\vec{k}\right| <\Lambda $ and $r_k^2=0$ otherwise.\ In the complex $\omega $%
-plane, the quark propagator has only on-shell poles at $\omega =\pm i\sqrt{%
\vec{k}^2+M^2}$.

However, when a 4-momentum regulator is used, the quark propagator acquires
extra poles, which also occur when proper-time regularization is used \cite
{Ripka95}. Such poles are unphysical in the sense that, taken seriously,
they lead to instabilities of the vacuum.\ Equivalently one can say that the
system behaves as if it was governed by a non-hermitian hamiltonian. This is
sometimes also expressed by saying that the theory becomes acausal. We
assign the cause of the instability discussed above to the existence of such
unphysical poles. When the 4-momentum cut-off is high (and $M_0/\Lambda $
correspondingly low) the effect of these unphysical poles is not felt.\ But
this is not the case in low energy hadronic physics where the parameter $%
M_0/\Lambda $ is in the range $0.4<M_0/\Lambda <0.8$.\ Furthermore, the
position (and therefore the effect) of the unphysical poles depends very
much on the shape of the regulator.

\begin{figure}[h]
\centerline{\resizebox*{12cm}{10cm}{\includegraphics{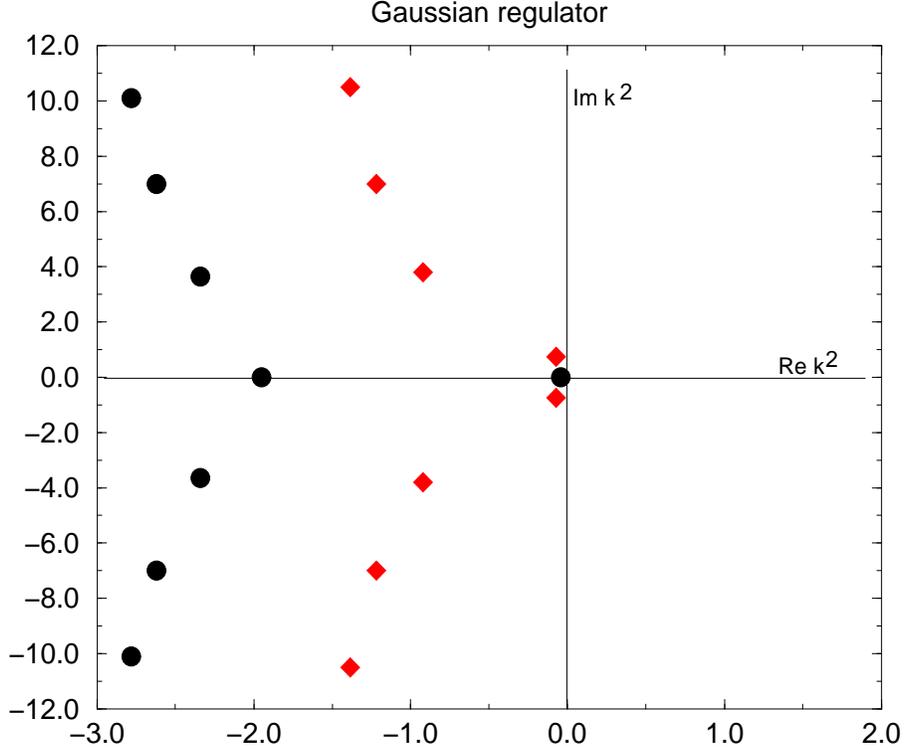}} }
\caption{The poles of the quark propagator with a gaussian regulator in the complex $k^2$ plane.\ The circles are for $M/\Lambda =0.2$ and the diamonds are for $M/\Lambda =0.8$.}
\label{epsfig59}
\end{figure}

A qualitative understanding of the difference between a sharp 4-momentum
cut-off and a soft gaussian regulator can be understood by comparing the
location of the poles of the quark propagator in the complex $k^2$ plane.\
When a gaussian regulator is used, the poles of the quark propagator occur
when: 
\begin{equation}
k^2+M^2e^{-\frac{2k^2}{\Lambda ^2}}=0
\end{equation}
Poles on the real axis of the complex $k^2$ plane occur only if $M^2<\frac{%
\Lambda ^2}2$. Otherwise and in addition, poles occur in the complex plane.
Figure \ref{epsfig59} shows the poles of the quark propagator in the two
cases $M/\Lambda =0.2$ (large cut-off) and $M/\Lambda =0.8$ (small cut-off).
The poles all lie to the left of the imaginary axis and they move closer to
it when the cut-off gets small. Table \ref{table:gausspole} gives the
residues of the poles.

\begin{table}[tbp] \centering%
$
\begin{tabular}{|c|c|c|c|c|c|c|c|}
\hline
$\frac M\Lambda $ & ${\rm Re}k^2$ & ${\rm Im}k^2$ & residue & $\frac
M\Lambda $ & ${\rm Re}k^2$ & ${\rm Im}k^2$ & residue \\ \hline
0.2 & -0.044 & 0 & $\allowbreak 1.1\,0$ & 0.8 & -0.072 & $\pm 0.74$ & $%
0.\,29\mp i\,\,0.\,51$ \\ \hline
0.2 & -1.941 & 0 & $-0.\,35$ & 0.8 & -0.906 & $\pm 3.81$ & $-0.01\,4\mp
i\,\,0.\,13$ \\ \hline
0.2 & -2.342 & $\pm 3.64$ & $-0.05\,5\mp i\,0.\,11$ & 0.8 & -1.202 & $\pm
6.98$ & $-0.007\,1\mp i\,\,0.071$ \\ \hline
0.2 & -2.607 & $\pm 6.89$ & $-0.02\,0\mp i\,0.06\,6$ & 0.8 & -1.386 & $\pm
10.14$ & $-0.004\,3\mp i\,\,0.04\,9$ \\ \hline
0.2 & -2.783 & $\pm 10.08$ & $-0.01\,1\mp i\,\,0.04\,7$ &  &  &  &  \\ \hline
\end{tabular}
$%
\caption{Position and residues of the poles of the quark propagator with a gaussian regulator.\ Values are given for $M/\Lambda =0.2$ and $0.8$
\label{table:gausspole}}%
\end{table}%

Consider next regularisation using a sharp 4-momentum cut-off.\ The
corresponding regulator does not have an analytic form, but we have checked
that very similar results are obtained with a Wood-Saxon shaped regulator $%
r_k^4=\frac 1{1+e^{\frac{k^2-\Lambda ^2}c}}$ which becomes equivalent to a
sharp cut-off when $c\rightarrow 0$.\ The poles of the quark propagator are
then the solution of the the equation: 
\begin{equation}
k^2+\frac{M^2}{1+e^{\frac{k^2-\Lambda ^2}c}}=0
\end{equation}
Setting $k^2=x+iy$, this equation decomposes into the two equations: 
\begin{eqnarray}
x+M^2\frac{1+e^{\frac{x-\Lambda ^2}c}\cos \frac yc}{\left( 1+e^{\frac{%
x-\Lambda ^2}c}\cos \frac yc\right) ^2+\left( e^{\frac{x-\Lambda ^2}c}\sin
\frac yc\right) ^2} &=&0 \\
y-M^2\frac{e^{\frac{x-\Lambda ^2}c}\sin \frac yc}{\left( 1+e^{\frac{%
x-\Lambda ^2}c}\cos \frac yc\right) ^2+\left( e^{\frac{x-\Lambda ^2}c}\sin
\frac yc\right) ^2} &=&0
\end{eqnarray}
One pole occurs on the negative real axis $\left( y=0\quad x\approx
-M^2\right) $ in the vicinity of $-M^2$. It is simple to see that the
complex poles all occur in the vicinity of the line $x=\Lambda $.\ The
imaginary part $y$ is then the solution of the equation: 
\begin{equation}
\frac y{2c}=\frac{M^2}{4c}\tan \frac y{2c}
\end{equation}
The spacing between the poles is thus close to $2\pi c$ so that they get
denser in number as $c\rightarrow 0$. In that limit, the continuum of poles
forms a cut which expresses the discontinuity of the regulator when a sharp
cut-off is used.\ A more exact numerical calculation of the position and
residues of the poles is given in the table \ref{table:sharppole} for the
case where $c=0.05$\thinspace $\Lambda ^2$ and $M/\Lambda =0.8$.

\begin{table}[tbp] \centering%
$
\begin{tabular}{|c|c|c|c|}
\hline
$M/\Lambda =0.8\quad c/\Lambda ^2=0.05\,$ & $x$ & $y$ & residue \\ \hline
$M=0.8\quad c=0.05$ & -0.64000 & 0 & 1 \\ \hline
$M=0.8\quad c=0.05$ & 1.02152 & $\pm 0.46354$ & $-0.02\,7646\pm 0.007\,3992i$
\\ \hline
$M=0.8\quad c=0.05$ & 1.01790 & $\pm 0.77472$ & $-0.02\,4736\pm 0.01\,1194i$
\\ \hline
$M=0.8\quad c=0.05$ & 1.01431 & $\pm 1.08764$ & $-0.02\,1203\pm 0.01\,3627i$
\\ \hline
$M=0.8\quad c=0.05$ & 1.01128 & $\pm 1.401618$ & $-0.01\,7739\pm 0.0\,148i$
\\ \hline
\end{tabular}
$%
\caption{Position and residues of the first few poles the quark propagator when a Wood-Saxon shaped regulator with $c=0.05$\thinspace $\Lambda ^2$ and $M/\Lambda =0.8$.
\label{table:sharppole}}%
\end{table}%

The unphysical poles produced by a soft gaussian regulator lie to the left
of the imaginary axis of the complex $k^2$ plane.\ They are therefore mostly
felt at low values of $k^2$ where phase space factors reduce their effect.\
The unphysical poles produced by a sharp cut-off lie close to the boundary $%
k^2=\Lambda ^2$ of high values of $k^2$ from which diverging quantities
derive most of their contribution. This explains qualitatively the
difference between the effect of the two regularizations on the effective
potential. It would be worth analyzing whether the instabilities, recently
heralded by Kleinert et al.\cite{Kleinert}, are not also artefacts of the
use of a sharp 4-momentum in conjunction with a low cut-off.

\section{Conclusions.}

Although the meson loop contributions do not modify appreciably the value of
the quark condensate, they do cause large quantum fluctuations of the quark
condensate. The Lorentz invariant regularization of the quark propagator,
used in conjunction with the relatively low cut-off values required in low
energy hadronic physics, makes the physical vacuum unstable against
distortions caused by the squared quark condensate.\ The instability depends
strongly on the shape of the regulator.\ It is only weakly felt when a soft
gaussian regulator is used but its effects are greatly enhanced in
calculations which use a sharp 4-momentum cut-off. Large errors can be made
if loop integrals are regularized after being derived from an unregularized
action, instead of including the regulator in the model action from the
outset. The ground state instability can be traced to unphysical poles of
the quark propagator which are introduced by the regulator.

\appendix\renewcommand{\theequation}{\Alph{section}.\arabic{equation}}%
\setcounter{equation}{0}

\section{Expressions for the classical action and the classical effective
action.}

\label{app:classact}

In the chiral limit, the value of the classical action at the stationaty
point is: 
\begin{equation}
I_{j,m=0}\left( M^2\right) -I\left( M_0^2\right)
\end{equation}
\begin{equation}
=\frac 124N_cN_f\frac \Omega {\left( 2\pi \right) ^4}\int d_4k\left( -\ln 
\frac{k^2+r_k^4M^2}{k^2+r_k^4M_0^2}+\,\,\frac{r_k^4M^2}{k^2+r_k^4M^2}-\frac{%
r_k^4M_0^2}{k^2+r_k^4M_0^2}\right)
\end{equation}
We measure all energies relative to the minimum $I\left( M_0^2\right) $ of
the classical action in the unconstrained system where $j=0$. In units of $%
\Omega \Lambda ^4$, the classical action depends only on the two variables $%
\frac M\Lambda $ and $\frac{M_0}\Lambda $.

\setcounter{equation}{0}

\section{Expressions for the meson propagators.}

\label{app:mesonprop}

\subsection{The meson propagators $K_{ab}^{-1}$ and the polarization
function $\Pi _a$ at a point $\varphi _a=\left( M,0,0,0\right) $.}

From the second order expansion of $-Tr\ln \left( -i\partial _\mu \gamma
_\mu +r\varphi _a\Gamma _ar\right) $ we obtain the following expression for
the polarization function: 
\begin{equation}
\Pi ^{\left( 2\right) }\left( \delta \varphi \right) =\frac 12Tr\frac
1{-i\partial _\mu \gamma _\mu +r^2M}r\delta \varphi _a\Gamma _ar\frac
1{-i\partial _\mu \gamma _\mu +r^2M}r\delta \varphi _a\Gamma _ar
\end{equation}
The calculation is standard.\ Taking traces over the Dirac and flavor
indices, and keeping track of the regulator, we obtain: 
\[
\Pi ^{\left( 2\right) }=4N_cN_f\frac 1{2\Omega }\sum_q\delta S\left(
q\right) \delta S\left( -q\right) \left( \frac 12q^2f_M^{22}\left( q\right)
+M^2\left( f_M^{26}\left( q\right) +f_M^{44}\left( q\right) \right)
-g_M\left( q\right) \right) 
\]
\begin{equation}
+4N_cN_f\frac 1{2\Omega }\sum_q\delta P_i\left( q\right) \delta P_i\left(
-q\right) \left( \frac 12q^2f_M^{22}\left( q\right) +M^2\left(
f_M^{26}\left( q\right) -f_M^{44}\left( q\right) \right) -g_M\left( q\right)
\right)
\end{equation}
where we wrote the fields $\varphi _a$ in terms of scalar and pseudoscalar
fields $S$ and $P_i:$%
\begin{equation}
\Gamma _a\varphi _a=S+i\gamma _5\tau _iP_i
\end{equation}
The Fourier transforms are defined to be: 
\begin{equation}
S\left( q\right) =\int d_4x\,\,e^{iq_\mu x_\mu }S\left( x\right) \quad \quad
\quad P_i\left( q\right) =\int d_4x\,\,e^{iq_\mu x_\mu }P_i\left( x\right)
\end{equation}
The function $f_M^{np}\left( q\right) $ is: 
\begin{equation}
f_M^{np}\left( q\right) =\frac{4\pi }{\left( 2\pi \right) ^4}\int_0^\infty
k^3dk\int_0^\pi d\alpha \sin {}^2\alpha \frac{r_{k_1}^nr_{k_2}^p}{\left(
k_1^2+r_{k_1}^4M^2\right) \left( k_2^2+r_{k_2}^4M^2\right) }  \label{fnpq}
\end{equation}
with $k_1=k-\frac q2$ and $k_2=k+\frac q2$. The function $g_M\left( q\right) 
$ is: 
\begin{equation}
g_M\left( q\right) =\frac{4\pi }{\left( 2\pi \right) ^4}\int_0^\infty
k^3dk\int_0^\pi d\alpha \sin {}^2\alpha \frac{r_{k_1}^2}{k_1^2+r_{k_1}^4M^2}%
r_{k_2}^2  \label{gmq}
\end{equation}
and we denote by $g_M$ the function $g_M\left( q=0\right) $.

It follows that the polarization function is diagonal in momentum space: 
\begin{equation}
\left\langle qa\left| \Pi \right| q^{\prime }b\right\rangle =\delta
_{ab}\delta _{qq^{\prime }}\Pi _a\left( q\right)
\end{equation}
where: 
\[
\Pi _{a=0}\left( q\right) \equiv \Pi _S\left( q\right) =4N_cN_f\left( \frac
12q^2f_M^{22}\left( q\right) +M^2\left( f_M^{26}\left( q\right)
+f_M^{44}\left( q\right) \right) -g_M\left( q\right) \right) 
\]
\begin{equation}
\Pi _{a=1,2,3}\left( q\right) \equiv \Pi _P\left( q\right) =4N_cN_f\left(
\frac 12q^2f_M^{22}\left( q\right) +M^2\left( f_M^{26}\left( q\right)
-f_M^{44}\left( q\right) \right) -g_M\left( q\right) \right)
\end{equation}
The inverse propagator matrix $K^{-1}$ is obtained by adding $-\left\langle
xa\left| \left( V-j\right) ^{-1}\right| yb\right\rangle $. We use the gap
equation (\ref{gap}) to write: 
\begin{equation}
-\left( V-j\right) ^{-1}=4N_cN_f\frac M{M-m}g_M
\end{equation}
so that: 
\[
K_S^{-1}\left( q\right) =4N_cN_f\left( \frac 12q^2f_M^{22}\left( q\right)
+M^2\left( f_M^{26}\left( q\right) +f_M^{44}\left( q\right) \right)
-g_M\left( q\right) +\frac M{M-m}g_M\left( 0\right) \right) 
\]
\begin{equation}
K_P^{-1}\left( q\right) =4N_cN_f\left( \frac 12q^2f_M^{22}\left( q\right)
+M^2\left( f_M^{26}\left( q\right) -f_M^{44}\left( q\right) \right)
-g_M\left( q\right) +\frac M{M-m}g_M\left( 0\right) \right)  \label{kaq}
\end{equation}

\vskip 0.5cm

{\Large \textbf{Acknowledgments.}}

\vskip 0.5 cm

The author wishes to thank W.Broniowski and B.Golli for numerous discussions, and 
J.Zinn-Justin and L.Pitaevskii for help in undestanding reference [11].

\bibliographystyle{unsrt}
\bibliography{njl}

\end{document}